\documentclass{emulateapj}

\shorttitle{S-star orbits}
\shortauthors{Merritt, Gualandris \& Mikkola}

\def\lap{\mathrel{\hbox{\rlap{\lower.55ex \hbox {$\sim$}}
        \kern-.3em \raise.4ex \hbox{$<$}}}}
\def\gap{\mathrel{\hbox{\rlap{\lower.55ex \hbox {$\sim$}}
        \kern-.3em \raise.4ex \hbox{$>$}}}}

\def\msun{M_\odot}

\begin{document}

\title{Explaining the Orbits of the Galactic Center S-Stars}

\author{David Merritt and Alessia Gualandris}
\affil{Department of Physics and Center for Computational Relativity and Gravitation, Rochester Institute of Technology, Rochester, NY 14623}
\author{Seppo Mikkola}
\affil{Tuorla Observatory, University of Turku, V\"ais\"al\"antie 20, Piikki\"o, Finland}

\begin{abstract}
The young stars near the supermassive black hole at the galactic center
follow orbits that are nearly random in orientation and that have
an approximately ``thermal'' distribution of eccentricities, $N(<e)\sim e^2$.
We show that both of these properties are a natural consequence
of a few million years' interaction with an intermediate-mass black hole 
(IBH), if the latter's orbit is mildly eccentric and if its
mass exceeds approximately 1500 solar masses.
Producing the most tightly-bound S-stars requires an IBH orbit
with periastron distance less than about $10$ mpc.
Our results provide support for a model in which the young stars
are carried to the galactic center while bound to an IBH,
and are consistent with the hypothesis that an IBH may still be orbiting
within the nuclear star cluster.
\end{abstract}


\section{Introduction}

Observations of the inner parsec of the Milky Way reveal two 
groups of massive young stars near the supermassive black hole (SBH).   
A group of roughly 40 stars move in approximately circular orbits that 
extend inward to a tenth of a parsec from the SBH 
\citep{PaumardApJ2006,LuApJ2008}.  
Another group of roughly 20 stars, the S-stars, follow eccentric 
orbits well inside the disk stars \citep{EckartMN1997}. 
The orbital periods of the S-stars are as short as 15 years and the 
orbits of these stars provide strong constraints on the mass and size 
of the central dark object \cite{GillessenApJ2008,GhezApJ2008}.

The presence of young stars so near the galactic center is a puzzle, since giant molecular clouds, which are the sites of star formation elsewhere in the galaxy, would be unable to collapse and fragment in the tidal field of the SBH \citep{MorrisApJ1993}.  
An orbiting gas cloud would instead be sheared into a set of rings; shocks between the rings would dissipate energy and the gas would settle into a thin disk.  If such a disk reaches a critical density, its self-gravity can overcome the tidal forces from the SBH allowing stars to form 
\citep{LevinApJ2003,NayakshinAA2005}.  
This model has been shown to successfully reproduce the observed properties of the young disk stars, including their top-heavy mass function and their mildly noncircular orbits \citep{BonnellScience2008}.  
But it fails to account for the S-stars, which move on eccentric, randomly-oriented orbits much closer to the SBH.

A number of models have been proposed to explain the S-stars but none is completely satisfactory \citep{AlexanderPhysRept2005,Paumard2008}. 
The S-stars could be old stars that migrated inward and were ``rejuvenated" by tidal heating or collisions with other stars.  
However their relatively normal spectra argue against such an exotic history 
\citep{Figer2008}.
Capture of young stars onto tightly-bound orbits via three-body exchange interactions involving compact remnants (e.g. stellar-mass black holes) may explain some of the S-stars \citep{AlexanderApJ2004}.
A similar model assumes that the S-stars were originally in binary systems; the more massive component of the binary was ejected during a close passage to the SBH, leaving the lower-mass star behind \citep{GouldApJ2003}.  
Both of these models require an ad hoc reservoir of new stars at large radii, as well as some mechanism for placing these stars onto plunging orbits soon after their birth, so that they can pass near the SBH where the chance of a three-body exchange is appreciable.  
The models also have difficulty reproducing the observed distribution of S-star orbital eccentricities.

An alternative scenario posutlates that the young stars formed in
a giant molecular cloud, far enough from the SBH that tidal forces 
did not preclude collapse and fragmentation \citep{GerhardApJ2001}.
The newly-formed star cluster then migrated inward via dynamical 
friction before tidal forces from the SBH dispersed it.  
The presence of an intermediate-mass black hole (IBH)
at the center of the cluster would assist in the transport
\cite{HansenApJ2003}; in the absence of the IBH, the cluster would be 
completely disrupted by tidal stresses at a distance of ~one parsec 
from the SBH \citep{ZwartApJ2003}.
This model requires a relatively high mass ($\gap 10^4 M_\odot$) 
and density for the cluster, although not much greater than what is 
observed in existing galactic center star clusters like the Arches and the 
Quintuplet \citep{Figer2008}.
While the timescale for formation of the IBH is uncertain, simulations
suggest that collisions between stars could form a massive remnant in a time
shorter than the time for cluster inspiral and dissolution 
\citep{ZwartApJ2002,FreitagMN2006}.
This model is also appealing since IBHs provide potential solutions to a 
number of other outstanding problems, including the origin of the 
hyper-velocity stars \citep{LevinApJ2006}, 
the structure of the stellar disks \citep{LevinApJ2005,BerukoffApJ2006},
and the growth of SBHs \cite{ZwartApJ2006}. 

Here we show that the infalling star cluster model can also reproduce 
the peculiar orbits of the S-stars.

\section{Assumptions}

We begin by summarizing two results from recent $N$-body simulations 
of the galactic center 
\citep{BaumgardtMN2006,MatsubayashiApJ2008,LockmannMN2008}.

\begin{itemize}
\item Inspiral of an IBH slows dramatically when it reaches a distance 
$\sim 10 (q/10^{-3})$ mpc from the SBH, where $q$ is the ratio of IBH to SBH 
masses; this distance is comparable to the sizes of the S-star orbits if 
$q \approx 10^{-3}$, i.e. if $M_{\rm IBH}\approx 10^{3.5}\msun$.  
Stalling occurs when the total binding energy in background (bulge) stars 
within the IBH orbit is comparable to that of the IBH itself; 
at this separation, most of the background stars that can exchange energy 
with the IBH are rapidly removed via the gravitational slingshot and the 
frictional force drops \citep{BBRNature1980}.

\item The orbit of the IBH is likely to be very eccentric at this late stage, 
$e \approx 0.5$ or greater, the exact value depending on the initial orbit of the star cluster containing the IBH and on the detailed history of IBH-star interactions after the cluster has been tidally removed. 
\end{itemize}

\noindent
We note that all of the $N$-body simulations cited above assumed an 
initially steep, $\rho\sim r^{-1.4}-r^{-1.75}$ density profile around the SBH.
In fact, there is evidence for a ``hole'' or dip in the density of 
the dominant, late-type (old) stellar population inside $\sim 0.5$ pc
\citep{FigerApJ2003,SchoedelAA2007,ZhuApJ2008}.
Accounting for the observed dip in the stellar densities would strengthen both
results summarized above: a lower density would decrease the dynamical
friction force on the IBH and lengthen the stalling phase;
and a shallower density profile is more conducive to eccentricity
increases \citep{GouldApJ2003}.

If sufficiently high eccentricities are reached, $e \gap 0.99$, 
gravitational radiation losses can shrink the SBH-IBH separation from 
$\sim 10$ mpc to coalescence in less than $10^8$ yr.  
In what follows, we assume that such extreme eccentricities are not attained  
and that the semi-major axis of the IBH orbit remains essentially 
unchanged for times comparable to S-star main-sequence lifetimes.  
However we argue below that these assumptions may not be strictly necessary
for producing the effects that we observe.

Prolonged gravitational interaction with the IBH can then scatter the young stars out of the thin disk into which they were originally deposited.  
The IBH acts on the stars in much the same way that Jupiter acts to scatter comets in the Solar system \cite{HansenApJ2003} -- 
with the important difference that Jupiter's orbit is nearly circular, 
while the IBH orbit is eccentric. 

\section{Methods} 

We carried out an extensive set of long-term $N$-body simulations to 
evaluate the effects of this interaction.  
Initial conditions consisted of a binary black hole representing the 
SBH-IBH pair and a set of 19, ten-solar-mass stars.  
Integration of each 21-body system was carried out using the computer 
program AR-CHAIN \citep{MikkolaAJ2008}, 
which incorporates an algorithmically regularized chain structure and 
time-transformed leapfrog to deal with near-collisions between the particles
\citep{MikkolaMN2006}.
Post-Newtonian corrections to the equations of motion were included 
up to order PN2.5.
Mikkola \& Merritt (2008) present a detailed description of the ARCHAIN 
algorithm as well as results of performance tests when the algorithm
is applied to galactic center problems very similar to the one treated here.

Initial conditions for the ``star'' particles were generated as follows.  
1. The position $r_{\rm apo}$ and velocity $v_{\rm apo}$ 
of the IBH at apastron were computed.  
2. A velocity of magnitude $F v_{\rm apo}$ was added with random direction 
to the IBH velocity.  
3. The Keplerian elements of the resulting orbit with respect to the SBH 
were computed, ignoring the presence of the IBH. 
4. A random value was assigned to the argument of the periastron, 
and a star was placed at a random phase on this orbit.  
This scheme produced an initial distribution of stars about the SBH 
that mimicked the phase-mixed distribution expected for a population 
of stars that were tidally stripped from the IBH \citep{BerukoffApJ2006}, 
with a (small) thickness determined by $F$ (see Fig.~1); 
all stars were initially orbiting in the same sense around the SBH.  

\begin{figure}
\includegraphics[angle=0.,width=0.46\textwidth]{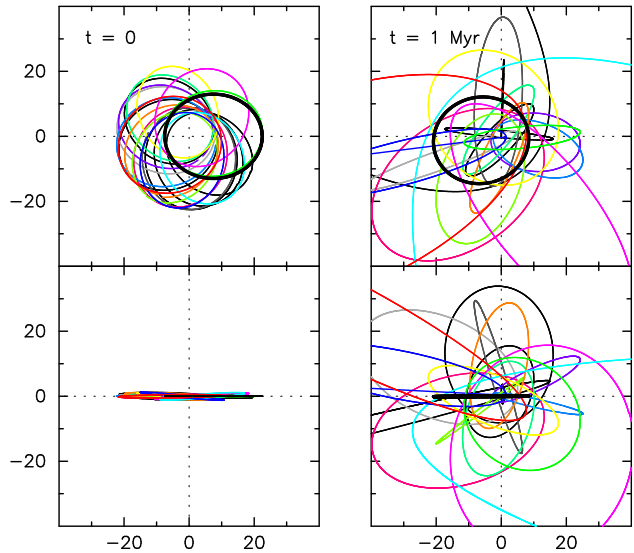}
\caption{
Initial (left) and final (right, after 1 Myr) orbits of stars in a 
simulation with IBH semi-major axis $a = 15$ mpc, 
eccentricity $e = 0.5$, IBH-SBH mass ratio $q = 0.001$, and $F = 0.2$.  
Top panels show the view looking perpendicular to the IBH orbital plane 
and bottom panels are from a vantage point lying in this plane.  
The IBH orbit is the heavy black curve in all panels; 
the unit of length is milliparsecs.  
The initially disk-like, co-rotating distribution of stars is converted, 
after 1 Myr, into an approximately isotropic distribution of orbits with 
a range of eccentricities, similar to what is observed for the S-stars.}
\end{figure}

In addition to $F$, the parameters that were varied were the mass ratio $q$, 
semi-major axis $a$, and eccentricity $e$ of the SBH-IBH binary.  
About 300 integrations were carried out, each for a time of $\sim 5$ Myr, 
based on an assumed SBH mass of $4.5\times 10^6 \msun$ 
and a distance to the Galactic center of 8.4 kpc.

\section{Results}

Figure~1 illustrates the evolution of the stellar orbits in one 
integration with $M_{\rm IBH} = 4500\msun$.  
In roughly one Myr, stars are scattered by the IBH (and occasionally 
by other stars) onto orbits with a wide range of eccentricities, 
semi-major axes and orientations.  

We measured the degree of randomness of the orbital planes using the 
Rayleigh (dipole) statistic $\cal{R}$ \citep{RayleighPhilMag1919} 
defined as the length of the resultant of the unit vectors 
$l_i, i = 1....N$, where $l_i$ is perpendicular to the orbital plane 
of the $i$th star.  
Initially, orbital planes are nearly aligned and ${\cal R}\approx N$, 
while for a random (isotropic) distribution ${\cal R}\approx N^{1/2}$.  
Figure~2 shows that ${\cal R}$ reaches values consistent with isotropy in a 
few Myr if the IBH mass is greater than $\sim 10^3\msun$ 
and if the eccentricity of the IBH orbit is $e \approx 0.5$ or greater.  
Eccentricity of the IBH orbit implies a semi-periodic forcing of the 
stars in a direction perpendicular to their initial orbital planes, 
allowing inclinations to be ``pumped up'' to large values after repeated 
encounters \citep{BinneyMN1981,ErwinApJ1999}.
When the orbit of the IBH is made nearly circular, $e\lap 0.2$, 
stellar orbital inclinations were found to remain nearly unchanged over 
these time scales.

\begin{figure}
\includegraphics[angle=0.,width=0.50\textwidth]{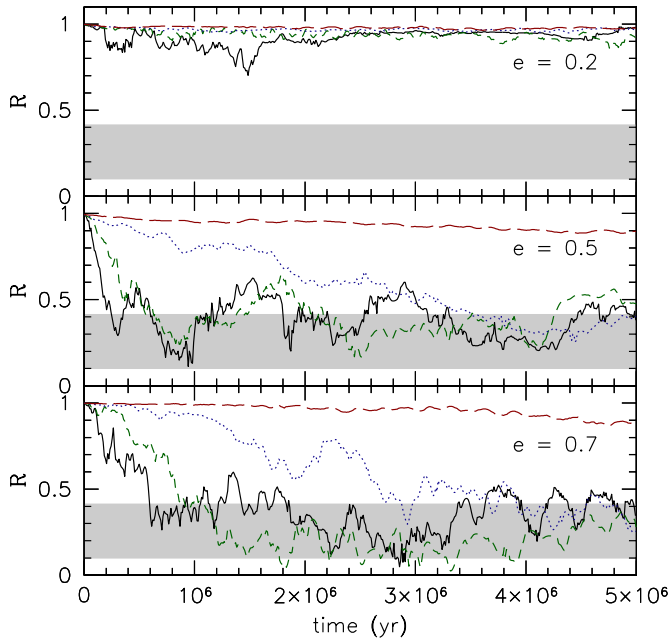}
\caption{Evolution of the Rayleigh parameter $R={\cal R}/N$ 
that measures the degree of randomness of the stellar orbital orientations, 
in 12 integrations, all with IBH semi-major axis $a = 30$ mpc.  
Black (solid) lines: $q = 10^{-3}$; green (dashed) lines: 
$q = 5\times 10^{-4}$; blue (dotted) lines: $q = 2.5\times 10^{-4}$; 
red (dashed) lines: $q = 10^{-4}$.  
The shaded regions show the $90\%$ confidence bands expected for a random, 
isotropic distribution of orbital orientations of 17 stars (the mean number 
of bound stars in the simulations). 
Gravitational perturbations from the IBH produce a nearly random 
distribution of orbital orientations after a few Myr, 
as long as the IBH mass and orbital eccentricity are sufficiently large.}
\end{figure}

\begin{figure}
\includegraphics[angle=0.,width=0.50\textwidth]{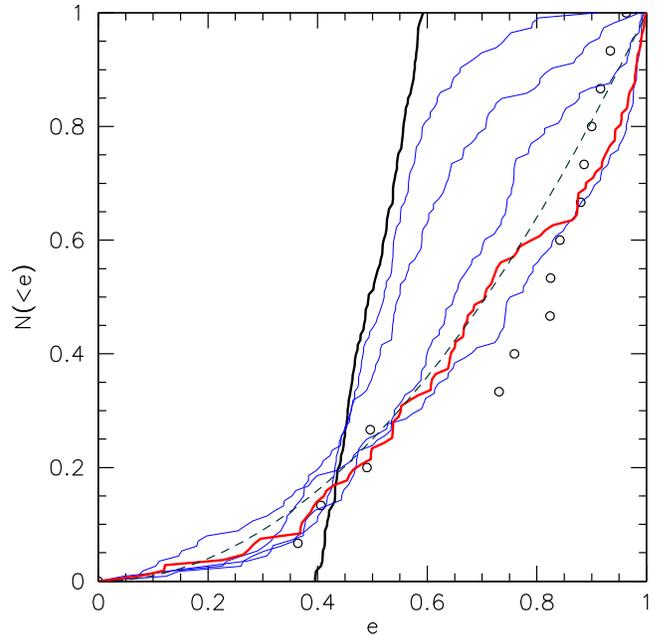}
\caption{Evolution of the distribution of stellar orbital eccentricities 
$e$ in a set of simulations with IBH orbital parameters $q = 5\times 10^{-4}$, 
$a = 15$ mpc, $e = 0.5$, $F = 0.2$.  
Each line is an average from six integrations with different random 
realizations of the initial conditions.  
Six times are shown: $t = 0$ (thick black curve), 
$t = (0.01, 0.02, 0.04, 0.2)$ Myr (thin blue lines), 
and $t = 1$ Myr (thick red line).  
The distribution is essentially unchanged at times greater than 1 Myr.  
Dashed line shows a ``thermal" distribution, $N\propto e^2$, 
and open circles are the eccentricity distribution observed for the 
S-stars \citep{GillessenApJ2008}.}
\end{figure}

The IBH also induces evolution in the eccentricities and energies 
(semi-major axes) of the stars. 
Eccentricities were found to tend toward a ``thermal'' distribution 
$N(<e) \sim e^2$  on a time scale of $\sim 0.1$ Myr for 
$q \gap 2.5\times 10^{-4}$ 
(Figure~3); this is similar to the distribution that is observed for the 
S-stars \citep{GillessenApJ2008}.   
In the case of orbital energies, perturbations from near encounters 
with the IBH can either increase or decrease them.  
Up-scattering in energy can continue until a star becomes unbound 
and escapes the system; this typically happened for at least one of the 
stars in each integration.  
Scattering to lower (more tightly bound) energies tended to be 
self-limiting: once a star is transferred to an orbit such that 
its apastron lies inside the periastron of the IBH, it is ``decoupled''
dynamically from the IBH and its energy tends to remain constant thereafter 
\citep{Wetherill1991}. 
Exceptions were observed only in the case $F = 0$; for such cold initial 
conditions, the stellar orbits remain highly correlated for long times 
encouraging strong interactions.  
Producing tightly-bound orbits like those of the innermost S-stars,
e.g. S2, therefore requires an IBH orbit with a periastron distance 
$a(1-e) \approx 10$ mpc.  
For IBH orbits satisfying this condition, we were able to reproduce 
approximately the full distribution of S-star semi-major axis lengths.  
However we do not consider this a crucial test since the observed sample 
is likely to be biased by radius-dependent selection effects.

We tested the extent to which the randomizing effects of the IBH
are helped by star-star scattering.
We carried out additional sets of integrations in which the masses
of the stars were decreased by factors of ten 
from $10\msun$ to $10^{-3}\msun$.
No significant dependence on stellar mass was observed, leading us
to conclude that interaction with the IBH is the dominant mechanism
responsible for the orbital evolution that we observe.

\section{Discussion}

We have found that the presence of an IBH orbiting within the nuclear
star cluster at the center of the Milky Way can efficiently randomize
the orbits of stars near the SBH, converting an initially thin, 
co-rotating disk into a nearly isotropic distribution of stars moving 
on eccentric orbits around the SBH.
Randomization of the orbital planes requires $\sim 1$ Myr if
the IBH mass exceeds $\sim 1500\msun$ and if the orbital eccentricity
$\sim 0.5$ or greater.
Evolution to a ``thermal'' eccentricity distribution occurs on an
even shorter time scale.
The final distribution of stellar semi-major axes depends on the
assumed size of the IBH orbit, but stars with apastron distances
as small as the {\it peri}astron distance of the IBH are naturally
produced.

Our simulations contribute only a small piece to the bigger unsolved
puzzle of the origin of the young stars at the galactic center.
If models for the genesis of the S-stars that invoke an inspiralling IBH are 
deemed otherwise viable,
our results show how the same models can also naturally reproduce
the random and eccentric character of the stellar orbits, 
and all in a time that is less than stellar evolutionary time scales 
-- thus providing an essentially complete explanation for the 
``paradox of youth" of the S-stars.

Although we chose parameters for the IBH such that gravitational 
radiation would not alter its orbit appreciably in 5 Myr, 
the rapidity with which the IBH modifies the stellar orbits in our 
simulations suggests that even an IBH on a decaying orbit might be 
able to randomize and ``thermalize" the S-star distribution before 
coalescing with the SBH.  
Thus, our model does not necessarily imply that an IBH is present, 
at the current time, within the S-star cluster.   
But if the IBH is there now, its presence might be detected in a 
number of ways:

1. The IBH will induce a motion of the SBH \citep{HansenApJ2003}.  
Upper limits on the astrometric wobble of the radio source Sgr A$^*$ 
are so far consistent with the presence of an IBH with mass and 
semi-major axis in the range considered here \citep{GillessenApJ2008}.

2. Stars can remain bound to the IBH if its Hill sphere is larger than 
its tidal disruption sphere; this condition is satisfied for SBH-IBH 
separations greater than $\sim 0.05$ mpc.  
The motion of a star bound to the IBH would be the superposition of a 
Keplerian ellipse around the SBH and an additional periodic component 
due to its motion around the IBH; the latter would have a velocity 
amplitude $\sim 0.1-10$ times the IBH orbital velocity and an orbital 
frequency from several hours to a few years, potentially accessible to 
astrometric monitoring.  

3. In favorable circumstances, a near encounter of the IBH with a star 
unbound to it could produce observable changes in the star's orbit 
over month- or year-long time scales. 

4. In our simulations with $q = 0.001$ and $e = (0.5,0.7)$, 
a fraction $13\%$ of the stars were ejected from the SBH-IBH system in 5 Myr.  
A star ejected at $\sim 10^3$ km s$^{-1}$ requires $\sim 10^2$ yr to move 
beyond 0.1 pc implying a probability $\sim 0.2 (N/10^4)$ of observing an 
escaping star at any given time in the Galactic center region, 
where $N$ is the number of stars subject to ejection.  
One S-star in fact appears to be on an escaping trajectory 
\citep{GillessenApJ2008}.

Finally we note that some models for IBH formation predict that a 
large number of IBHs might co-exist in the galactic center region 
\citep{ZwartApJ2006}.
If so, the rate of orbital evolution of the young stars might be 
even higher than what we observe in our simulations with a single IBH.

\acknowledgments
D. M. and A. G. were supported by grants AST-0807910 (NSF) and 
NNX07AH15G (NASA). 
We thank Don Figer and Milos Milosavljevic for comments that
improved the manuscript.

{}

\end{document}